\documentclass[11pt]{article}
\usepackage{setspace}
\usepackage{pstricks}
\usepackage{graphicx}
\usepackage[left=4cm,top=3cm,bottom=4cm,right=4cm]{geometry}
\usepackage{amsthm}
\usepackage{amssymb}
\usepackage{amsmath}
\usepackage{stmaryrd}
\usepackage{xcolor}
\usepackage{url}
\theoremstyle{definition}

\newtheorem*{prf}{Proof}

\newtheorem{thm}{Theorem}
\newtheorem{defn}{Definition}
\newtheorem{lem}{Lemma}
\newtheorem{prop}{Proposition}
\newtheorem{col}{Corollary}

\title{On the Cardinality of Future Worldlines in Discrete Spacetime Structures}

\author{Ahmet \c{C}evik\footnote{Gendarmerie and Coast Guard Academy, Ankara, Turkey. E-mail: a.cevik@hotmail.com}\hspace{0.5cm} Zeki Seskir\footnote{Middle East Technical University, Department of Physics, Ankara, Turkey. Email: zseskir@metu.edu.tr}}
\date{ }      
\begin{document}
\maketitle
\begin{abstract}
We give an analysis over a variation of causal sets where the light cone of an event is represented by finitely branching trees with respect to any given arbitrary dynamics. We argue through basic topological properties of Cantor space that under certain assumptions about the universe, spacetime structure and causation, given any event $x$, the number of all possible future worldlines of $x$ within the many-worlds interpretation is uncountable. However, if all worldlines extending the event $x$ are `eventually deterministic', then the cardinality of the set of future worldlines with respect to $x$ is exactly $\aleph_0$, i.e., countably infinite. We also observe that if there are countably many future worldlines with respect to $x$, then at least one of them must be necessarily `decidable' in the sense that there is an algorithm which determines whether or not any given event belongs to that worldline. We then show that if there are only finitely many worldlines in the future of an event $x$, then they are all decidable. We finally point out the fact that there can be only countably many terminating worldlines.
\end{abstract}

\noindent {\small {\bf Keywords} Multiverse, discrete spacetime, mathematical cosmology, determinism, causal sets, closed timelike curves, effectively closed sets, Cantor space, computable trees.}

\vspace{0.5cm}
The many-worlds interpretation (MWI) of quantum mechanics has been a topic of interest since the inception of the idea in Hugh Everett's doctoral thesis and a following publication \cite{wheeler1957}. His work was carried under the supervision of John Archibald Wheeler, hence it is usually referred to as the {\em Everett-Wheeler interpretation} of quantum mechanics. The original article has more than 1400 citations to date, and it has been depicted in the pop culture with many cultural references to the idea of parallel universes. Furthermore, with the emergence of quantum computing as a field of importance, and through David Deutsch's, one of the early leading figures in quantum computing, open adoption of this interpretation in his book \cite{deutsch97}, the topic has been revitalized. Although the exact ratio of physicists considering this as their main choice of interpretation of quantum mechanics is unknown and debated in many studies including \cite{zeilenger2013} and \cite{norsen2013}, it is considered as an alternative to the generally accepted Copenhagen interpretation (CI). 

There are different CI and MWI formulations, but to crudely summarize the main difference between these two interpretations one can focus on the problem of measurement. On CI, the wavefunction describing the quantum state of a physical object `collapses' at the moment of measurement into one of the possible states. The probability distribution before the point of measurement becomes irrelevant, and the physical reality only consists of the outcome of this measurement. Conversely, on MWI, the point of measurement does not cause an irreversible effect, all the possibilities before continue to exist in orthogonal worlds.

One intriguing question concerns exactly how many worlds there are in this interpretation. The question was also asked in \cite{healey84} and \cite{linde2010}. One way of investigating this is to compute it numerically. We are rather interested in counting `in the limit', where the universe expands indefinitely. For example, if we assume that {\em all} possible worldlines have an ending, then the number of possible worldlines is finite.\footnote{For details, see the discussion after Corollary \ref{col:corollary2}.} On the other hand, if we do not have any restriction on the lifespan of worldlines, then clearly there may be infinitely many of them. However, as Georg Cantor pointed out nearly 150 years ago, in his \cite{cantor1872} and other works, there are different sizes of infinities. So it becomes a natural question to ask exactly how many possible worlds there are if there exist infinitely many of them. In this study we try to devise a rigorous account of how many worlds there might be in different scenarios in the MWI, under reasonable assumptions about spacetime, causality, and events. 

It is a major and longstanding debate in the philosophy of physics whether spacetime fabric admits a discrete or continuous structure. There are various physical theories like general relativity that support the continuous view based on Lorentzian manifolds, as well as theories that rely on a discrete space conception such as the causal set account of quantum gravity. This dichotomy also determines how we perceive events and the way they occur in spacetime. In this work we take the discrete route and represent events as discrete indivisible spacetime entities. This is also the route taken in what is known today as {\em causal set theory}, introduced in \cite{Bombelli1987}, and studied by many researchers, including works in \cite{Dowker2006}, \cite{BombelliLeeMeyerSorkin1987}, \cite{Meyer1988}, \cite{Reid2001}, \cite{Surya2019}, \cite{HuggettWutrich}, \cite{Dribus2017}. A {\em causal set} $C$ is a partially ordered set with a binary order relation $\leq$ such that for every $x,y,z\in C$, it satisfies the following conditions.
\begin{enumerate}
    \item[(i)] Reflexivity: $x\leq x$
    \item[(ii)] Transitivity: if $x\leq y$ and $y\leq z$, then $x\leq z$.
    \item[(iii)] Anti-symmetric (acyclicity): if $x\leq y$ and $y\leq x$, then $x=y$.\footnote{Without (iii) we get a pre-order. In fact, (iii) avoids cycles, that is, closed time-like curves, and as pointed out in \cite{HuggettWutrich}, `events on a causal loop would therefore not be distinct at all and the theory lacks the resources the distinguish between a single event and a causal loop' (p. 11).}
    \item[(iv)] Local finiteness: $|\{z:x\leq z\leq y\}|<n$ for some natural number $n$, where for a set $S$, $|S|$ denotes the number of elements or the {\em cardinality} of $S$.
    \item[(v)] $C$ is countable.\footnote{This condition is usually omitted, but it is stated in \cite{HuggettWutrich}.}
\end{enumerate}
Causal sets can thus be represented by acyclic directed graphs or by their Hasse diagrams. Essentially, causal sets replace continuum spacetime concept due to a theorem by Malament in \cite{Malament1997}: `The metric of a globally hyperbolic spacetime can be reconstructed uniquely from its causal relations up to a conformal factor' (p. 1399).\footnote{We shall call this {\em Malament's theorem} for future reference.} 

It is known due to \cite{DowkerZalal2017} that `the number of elements in any portion of the causal set manifests itself, on average, as spacetime volume of the corresponding region of the approximating continuum' (p. 247). This leads to Sorkin's slogan that `order + number = geometry'. According to Bombelli {\em et al} \cite{BombelliLeeMeyerSorkin1987}, (1) quantum gravity is a quantum theory of causal sets; (2) a continuum spacetime $(M,g)$ is an approximation of an underlying causal set $C$, where the order of the elements of $C$ approximates to the causal order in a continuum spacetime, and the number of elements in a region of causal set approximates to continuum spacetime volume. This is where the dynamics of causal sets comes into play. However as we said earlier, the analysis given in this paper is independent of any dynamics.

Most of the works in causal set theory, such as \cite{Sorkin2005DiscreteGravity}, \cite{Yamamoto1989}, \cite{Kempf2010}, \cite{Bombelli2009} introduce the discreteness property while preserving Lorentz invariance.\footnote{For details, we refer the reader to \cite{Hu2013PhD}, particularly p. 20.} We consider a rather `digitalized' version of causal sets so as to discretize the light cone where events occur in least possible time ticks. In this regard, it should be noted that, although our construct has many of the features of causal sets, such as a discrete partial order with a form of light cone structure, it diverges from the standard formulation of causal sets, for example, we don't assume a particular relationship between causal sets and spacetime manifolds other than the assumptions we introduce in Section 1, as our analysis is independent of any dynamics of causal sets. We will only work with the worldlines of light cones. Since our analysis mainly concerns the properties of infinite worldlines of a light cone, the results that we will give are independent of the dynamics of causal sets. This is due to the fact that we are only concerned with the {\em limit} case of a grown causal structure, whereas the concept of introducing dynamics only makes sense to differentiate a growing causal structure at different fixed stages. So one may just take, for instance, the classical sequential growth discussed in \cite{Rideout1999}, however it does not really matter what type of dynamics is applied. For further information on causal set dynamics, we refer the reader to \cite{Wallden2013}. Our assumptions also show some similarities with the {\em branching space-time}, which originally goes back to Belnap \cite{Belnap1992}, who proposed a branching space-time structure as a `simple blend of relativity and indeterminism'.\footnote{See \cite{BelnapMullerPlacek2021} for a revised version of branching space-time.} The project was criticized in \cite{Earman2008} where the authors suggested to prune some branches from branching space-time. In \cite{Muller2013}, M\"{u}ller then introduced a differential-geometrical version of branching space-times as a generalized non-Hausdorff manifold. In \cite{Vanchurin2015}, Vanchurin used discrete tree-like structures for the space-time generated by eternal inflation to investigate the measure problem.

Consequently with our version of causal sets, under appropriate assumptions on the properties of the universe, one can describe light cones as finitely branching tree structures and define worldlines as possibly infinite paths on such trees. The set of all infinite paths on a computable tree will be viewed as an effectively closed subset of $2^\mathbb{N}$ Cantor space, where $\mathbb{N}$ denotes the set $\{0,1,2,\ldots\}$ of natural numbers.\footnote{The word `effective' means `computable'. Secondly, we will in fact use the space $n^\mathbb{N}$ for arbitrary $n\in\mathbb{N}$, which we shall refer to it with the same name since $2^\mathbb{N}$ and $n^\mathbb{N}$ are topologically the same.} Using some facts about effectively closed subsets of Cantor space, we observe that for any event $x$, if all worldlines extending $x$ are `eventually deterministic', in the sense that there exists an event above which there is no more split in the light cone tree of $x$, then the number of possible worldlines with respect to $x$ is exactly $\aleph_0$, i.e., the cardinality of $\mathbb{N}$. Furthermore, if there are countably many possible worldlines with respect to $x$, then at least one of them must be decidable in the sense that there is an effective method, i.e. an algorithm, which determines whether or not any given event belongs to the given worldline. 

The paper is organised as follows. In Section \ref{sec:basicassumptions}, we start by giving some basic assumptions about the universe, about causality, and events. In Section \ref{sec:worldlinesasreals} we translate these physical concepts into mathematical domain by viewing light cones of events as finitely branching trees. Consider the following statements:

\begin{enumerate}
    \item [(i)] All worldlines extending a given event $x$ are `eventually deterministic'.\footnote{We will define what we mean by eventually deterministic in Section \ref{sec:alternateuniverses}.}
    \item[(ii)] The number of possible worldlines with respect to $x$ is exactly $\aleph_0$. 
    \item[(iii)] There exists a `decidable' worldline.
\end{enumerate}

We show that (i) is a sufficient condition for (ii). We also show that (ii) implies (iii), and thus (i) implies (iii). 

In other words, we argue that given any event $x$, if all worldlines extending the event $x$ are `eventually deterministic', then there are exactly $\aleph_0$ possible future worldlines with respect to $x$. We then observe that if there are countably many possible worldlines with respect to $x$, then at least one of these worldlines must be necessarily `decidable' in the sense that there is an algorithm which determines whether or not any given event belongs to the given worldline. We also give some corollaries, and discuss about the conclusions in the final section.

\section{Basic Assumptions}\label{sec:basicassumptions}

This section lays out the assumptions about the structure and the property of the physical universe and of the concept of causality in order to prepare us to work in the mathematical domain. The reason one must articulate these hypotheses is that it is the only way to move the subject matter into the mathematical domain so as to develop some logico-mathematical results. Although the assumptions given in this paper may be disputable, the results we show will be rigorous and clear. One can discuss whether any or all of these hypotheses are not representative of the physical reality that we are living in, however we tried to keep these assumptions on the most fundamental level and in accordance with up-to-date cosmological understanding to the best of our abilities.

We begin by giving the following assumptions about the universe and then move onto giving some axioms concerning causality and events. In the final part, we give our definition of worldline. The order between these items is irrelevant. However, we shall list them starting from the simplest assumption and get more specific as we move on. 
\vspace{0.5cm}

\noindent {\bf I. Assumptions about the universe}

\begin{enumerate}

    \item [(a)] The universe is expanding indefinitely.\footnote{Whether the universe has a beginning or not will not be relevant to our analysis and will not affect the results for reasons that will become clear later. However, since we will be concerned with spacetime light cones of events in its general form---particularly the future light cone of events---we may assume without loss of generality that the past and future cones are both {\em unbounded}, and so we will make our analysis under the assumption that the future light cone of any event is potentially unbounded. Eternal past can be assumed in an oscillating universe model. Nevertheless, we will be primarily concerned with the future light cone of events only.}
    \item[(b)] (i) There are finitely many atoms in the observable universe at any fixed moment. (ii) Every atom consists of at most finitely many subatomic particles.
    \item[(c)] By (a) and (b), it follows that the observable universe is finite.
    \item[(d)] Spacetime is not dense.\footnote{In fact, whether spacetime is discrete or continuous is not a settled question in physics as both are consistent with different physical theories. See \cite{Forrest1995} for a discussion.} That is, for any two distinct spacetime points $x$ and $y$, there are at most finitely many points in-between $x$ and $y$ in the Euclidean sense. That is, 
    \[
    |\{z:bet(z,x,y)\}|<\aleph_0,
    \]
    where $bet(z,x,y)$ denotes that the point $z$ is between $x$ and $y$. 
    \item[(e)] The {\em no-signalling principle}, as given in \cite{masanes2006}, is valid. That is, events are bounded by the speed of light in their future light cones. 
\end{enumerate}

\noindent {\bf II. Assumptions about causality and events}
\begin{enumerate}
    \item [(a)] Let $x$ and $y$ be two events.\footnote{By {\em event} we mean an instantaneous situation or action that is associated with a point in spacetime. Events are primitive objects of the domain of discourse in any theory of causal sets.} We let $x\leq y$ mean ``event $x$ is a cause of event $y$" (or ``event $y$ is an effect of $x$). We let $x<y$ mean $x\leq y$ and $x\neq y$. Any given event $x$ defines a double light cone consisting of lower and upper parts, where the {\em lower cone} is defined as the collection of all possible past events that end up with $x$. That is, the lower cone of $x$ is
    \[
    \{y:y\leq x\}.
    \]
    The {\em upper cone} represents all possible future events that stems from $x$, which is defined as the set
    \[
    \{y:x\leq y\}.
    \]
    It is mathematically natural to assume, given two events $x$ and $y$, if their upper (or lower) light cones are equal to each other, then $x=y$. That is, the past or the future of every distinct event $x$ gives a unique set of events.\footnote{Chronologically speaking, these are called {\em future and past distinguishing spacetimes} for which Malament's theorem proves the equality of the existence of a chronological bijection and the existence of conformal isometry. Levichev (1987) \cite{Levichev1987} then showed that a causal bijection implies a chronological bijection and hence Malament's theorem can be generalized to causal bijections.}

    \item[(b)] Causality is transitive. That is, if $x\leq y$ and $y\leq z$, then $x\leq z$. 
    
    \item[(c)] Causality is not dense, i.e., if $x\leq y$ then there do not exist infinitely many events occurring between the event $x$ and the event $y$.\footnote{This is also known as the {\em locally finiteness} condition in causal sets.}
    
    \item[(d)] We say that two events $x$ and $y$ are {\em incomparable} (or {\em spacelike separated}) if neither $x\leq y$ nor $y\leq x$. Two incomparable events $x$ and $y$ cannot be seen as a single event. Similarly, no event can be decomposed into smaller incomparable sub-events. In other words, all events are `atomic' (see Section \ref{sec:translation} for details).
    
    \item[(e)] For every event $x$, since there are finitely many atoms---by I(b) and I(d)---there can only be finitely many events $y_1,\ldots,y_n$ such that $y_i$ is the {\em immediate successor} (or {\em immediate effect}) of $x$, meaning that $x<a<y_i$ for no event $a$ (See Section \ref{sec:translation} for details).
    
    \item[(f)] A {\em closed timelike curve} is a sequence of events that happen between two distinct events $x$ and $y$ such that $x\leq y$ and $y\leq x$. There exist {\em closed timelike curves} iff the {\em anti-symmetric property} does not hold for causality. In other words, there are closed timelike curves iff it is not the case that $x=y$ whenever $x\leq y$ and $y\leq x$.\footnote{In most theories of causal sets, the anti-symmetric property is implicitly assumed. We follow the same tradition.}  
    
    \item[(g)] It follows from I(b), II(c), and II(e) that given any event $x$, the collection of all future events that stem from $x$ forms a subset of $n^{<\mathbb{N}}$, for some $n\in\mathbb{N}$ (see Section \ref{sec:translation} for details).
    
    \item[(h)] The collection of all events which end up with $x$ gives the lower cone of $x$ and it can be similarly represented by a subset of $n^{<\mathbb{N}}$, for some $n\in\mathbb{N}$ (see Section \ref{sec:translation} for details).

\end{enumerate}

\noindent {\bf III. Assumptions on worldlines}
\begin{enumerate}
    \item[(a)] Let $x$ be an event. A {\em worldline} of $x$ is a linearly ordered set 
    \[
    \{\ldots,p_2(x),p_1(x),x,f_1(x),f_2(x),\ldots\}
    \]
where each $p_i(x)$ denotes a single past event of $x$ such that $p_i(x)\leq x$ and that $p_{i+1}(x)\leq p_i(x)$ for every $i$, whereas each $f_i(x)$ denotes a single future event satisfying that $x\leq f_i(x)$ and $f_i(x)\leq f_{i+1}(x)$ for every $i$. A {\em future worldline} of $x$ is the set of events that causally proceed $x$, that is,
\[
\{x,f_1(x),f_2(x),\ldots\}.
\]
A {\em terminating worldline} is a worldline in which there exists an event $e$ such that $e\leq e'$ for no event $e'$. 
\end{enumerate}

We shall now discuss about the consequences of these basic assumptions and also argue what worldlines actually correspond to in our model. 

\section{Worldlines as real numbers}\label{sec:worldlinesasreals}

We first give some basic notions and facts about effectively closed subsets of Cantor space and computable trees. These notions can be found in more detail in \cite{Diamondstone}, and \cite{CenzerPi01classes}. We then give some results and corollaries that will apply to causal sets in this representation. After giving the results, we elaborate more on the basic assumptions given in Section \ref{sec:basicassumptions} to interpret these physical notions in the mathematical domain.

\subsection{Computable subsets of Cantor space}
\label{sec:cantorspace}

When we say a {\em computable set} we mean a set $A\subseteq\mathbb{N}$ for which there is an algorithm such that for any given natural number $n$, the algorithm can decide whether or not $n\in A$.\footnote{We may use the terms {\em computable} and {\em decidable} interchangably depending on the context.} Similarly, an $n$-ary relation $R$ is computable if it is computable as a set, that is, if for any given $n$-tuple $(x_1,\ldots,x_n)$, there is an algorithm which can decide whether or not $(x_1,\ldots,x_n)$ is in $R$. A set $A$ is {\em countable} if it is either finite or there is a one-to-one correspondence between $A$ and $\mathbb{N}$. If a countable set is infinite, then we call it a {\em countably infinite} set. If a set is not countable, we call it {\em uncountable}. {\em Cantor's theorem} says that the cardinality of a set $A$ is strictly smaller than the cardinality of the set of all subsets of $A$.

Any countable set is in one-to-one correspondence with the set of natural numbers under an appropriate pairing function. The definition of a computable subset of naturals then generalizes to any countable set under a suitable pairing function. Let us now start by giving the notation for strings and trees. 
\vspace{0.5cm}

\noindent{\bf Strings.}
A {\em string} is a finite sequence of symbols. We denote finite strings with lowercase Greek letters like $\sigma,\tau,\eta,\rho,\pi,\upsilon$. We let $\sigma*\tau$ denote the concatenation of $\sigma$ followed by $\tau$.\footnote{Note that the concatenation operation is not commutative.} We let $\sigma\subseteq\tau$ denote that $\sigma$ is an initial segment of $\tau$. For example, $1001\subseteq 10011$. We say a string $\sigma$ is {\em incompatible with $\tau$} if neither $\sigma\subseteq\tau$ nor $\tau\subseteq\sigma$. Otherwise we say that $\sigma$ is {\em compatible} with $\tau$. Similarly, we say that $\sigma$ is an {\em extension} of $\tau$ if $\tau\subseteq\sigma$. We let $|\sigma|$ denote the {\em length} of $\sigma$, i.e., the number of characters in $\sigma$. The {\em empty string} is the unique string of length $0$. For any string $\sigma$, the concatenation of $\sigma$ with the empty string is equal to $\sigma$.

Without loss of generality we may use the ordinal number notation and identify the ordinal 2, for instance, with the set of smaller ordinals $\{0,1\}$. A set $A\subseteq\mathbb{N}$ can be identified with its characteristic function $f:\mathbb{N}\rightarrow\{0,1\}$ such that $f(n)=1$ if $n\in A$, $f(n)=0$ otherwise. We represent the set of these functions as $2^\mathbb{N}$. This can be generalized to $n^\mathbb{N}$ for any natural number $n$. A set $A\subseteq\mathbb{N}$ codes an infinite 0-1 binary string if we take its characteristic function as a sequence of 0's and 1's so that if $n\in A$, then the $(n+1)$-th digit of the characteristic sequence of $A$ is 1, and it is 0 otherwise. So when we write $\sigma\subseteq A$ we mean $\sigma$ is an initial segment of $A$ as a sequence. 

\vspace{0.5cm}

\noindent {\bf Trees and $\Pi^0_1$ classes.}
A set $T$ of strings is {\em downward closed} if $\tau\in T$ whenever $\sigma\in T$ and $\tau\subseteq\sigma$. A {\em tree} is a downward closed set of strings. For $n\in\mathbb{N}$, an {\em $n$-ary tree} is a tree $T$ with at most $n$ many branchings for each element in $T$. We say that a set $A$ is an infinite {\em path} on $T$ if there exist infinitely many $\sigma$ in $T$ such that $\sigma\subseteq A$. A set is a finite path if there are finitely many such $\sigma$'s. So if $A$ is an infinite path on $T$, then every initial segment of $A$ is in $T$. For $n\in\mathbb{N}$, if $T\subseteq n^{<\mathbb{N}}$ is a tree, then the set of infinite paths on $T$ is defined as
\begin{center}
$\left[T\right]=\{A:\forall n(A\upharpoonright n\in T)\}$,
\end{center}
where $A\upharpoonright n$ denotes the initial segment of the path $A$ up to length $n$. The notation $A\in[T]$ means $A$ is an infinite path on $T$.

We say that a string $\sigma\in T$ is {\em infinitely extendible} in $T$ if there exists some $A\supseteq\sigma$ such that $A\in[T]$. A tree $T$ is {\em perfect} if every infinitely extendible string in $T$ has at least two incompatible extensions in $T$. If $\sigma,\tau\in T$ and $\sigma\subseteq\tau$ and there does not exist $\sigma'$ with $\sigma\subseteq\sigma'\subseteq\tau$ then we say that $\tau$ is an {\em immediate successor} of $\sigma$ in $T$ and $\sigma$ is the {\em immediate predecessor} of $\tau$ in $T$. Given the definitions above, we will view each event as a string, that is, an element of a tree, and view each worldline as a finite or infinite path. We give a discussion about this more in Section \ref{sec:translation}.

\begin{defn}
We say that a tree $T$ of strings is {\em computable} if for any string $\sigma$, there is an algorithm which decides whether or not $\sigma\in T$.
\end{defn}

\begin{defn}
$\mathcal{A}\subseteq n^\mathbb{N}$ is called a {\em $\Pi^0_1$ class} (or {\em effectively closed set }) if there exists a computable relation $\varphi(n,X)$ such that $X\in \mathcal{A}$ if and only if $\forall n \ \varphi(n,X)$, where $n$ ranges over $\mathbb{N}$ and $X$ ranges over $n^\mathbb{N}$.
\end{defn}

\noindent From the definitions it is easy to see the following.

\begin{thm}
Suppose that $\mathcal{A}\subseteq n^\mathbb{N}$ for some natural number $n$. The following are known to be equivalent.
\begin{enumerate}
\item[(i)] $\mathcal{A}=\left[T\right]$ for some computable tree $T$.
\item[(ii)] $\mathcal{A}$ is effectively closed.
\end{enumerate}
\end{thm}

For a proof of this theorem, we refer the reader to \cite{Diamondstone} (pp. 129-130), as well as \cite{Nies2009} (p. 53).

Since we work in Cantor space, we shall mention the compactness property of it. Although compactness can be provided by various forms, the most well known form is given by {\em K\"{o}nig's lemma} \cite{Konig}.

\begin{lem}[K\"{o}nig's lemma]
If $T$ is a finitely branching infinite tree, then $T$ has an infinite path.
\end{lem}
\begin{prf}
Suppose we are given a finitely branching infinite tree $T$. We define a path $A=\bigcup_{n\in\mathbb{N}}\sigma_n$ on $T$ by induction on $n$. First we may define $\sigma_0$ to be the empty string, i.e., the root of $T$. Given $\sigma_n\in T$ for which there are infinitely many extensions in $T$, we define $\sigma_{n+1}$ to be an immediate successor of $\sigma_n$ in $T$ such that $\sigma_{n+1}$ has infinitely many extensions in $T$. Such $\sigma_{n+1}$ exists because $\sigma_n$ has infinitely many extensions in $T$, but only finitely many immediate successors since $T$ is finitely branching. Hence, at least one of the immediate successors must have infinitely many extensions in $T$.\qed
\end{prf}

We shall now give some topological properties about finitely branching trees.

\begin{defn}
Let $n\in\mathbb{N}$ and let $T\subseteq n^{<\mathbb{N}}$ be a tree.
\begin{enumerate}
\item[(i)] For any given $\sigma\in T$, we let $T_\sigma$ be the {\em subtree of nodes compatible with $\sigma$} and be defined as
\begin{center}
$T_\sigma=\{\tau\in T:\sigma\textrm{ is compatible with }\tau\}$.
\end{center}
\item[(ii)] A path $A\in\left[T\right]$ is said to be {\em isolated} if there exists a string $\sigma\in T$ such that $\left[T_\sigma\right]=\{A\}$, in which case we say that $\sigma$ {\em isolates} $A$. Otherwise $A$ is called a {\em limit point}.
\end{enumerate}

\end{defn}

Note that when $\sigma$ isolates $A$, there are no incompatible infinite extensions of $\sigma$ in $T$. We remark the following proposition for the next subsection, although the proof is immediate from Cantor's theorem. 

\begin{prop}\label{prop:perfect}
Let $T$ be a perfect tree. Then $[T]$ is uncountable. 
\end{prop}
\begin{prf}
If $T$ is a perfect tree, then by definition every infinitely extendible string in $T$ has at least two incompatible extensions in $T$. Thus, the collection of infinite paths on $T$ has at least $2^{\aleph_0}$ members, which is known to be equal to the cardinality of the set of real numbers, which is uncountable by Cantor's theorem.\qed
\end{prf}

\subsection{Possible worlds in causal sets}
\label{sec:alternateuniverses}

The axioms of causal sets given in Section 1 describe a discrete spacetime structure on which a partial order is defined by causality. We said that each event is represented by an element of a tree. So if $\sigma$ and $\tau$ are two strings in $T$, and that $\sigma\subseteq\tau$, this means that the event denoted $\sigma$ is a cause of the event denoted by $\tau$ in the causal structure. Given any event $x$, the future light cone of $x$ is represented by a full $n$-ary tree $T$. So $T$ defines all possible events caused by $x$. Worldlines are defined by the collection of all paths on $T$, and a full $n$-ary tree is necessarily an uncountable set since it is perfect.\footnote{For further discussion on how we define a tree for a light cone of an event, see Elaboration of III(a) in Section \ref{sec:translation}.} But we ask under what conditions there could be less worldlines defined on $T$. Clearly, if all possible worldlines are terminating, then there must be only finitely many of them since there will always be an event in every worldline after which no event occurs.\footnote{See the discussion at the end of Section \ref{sec:alternateuniverses} for more details.} A more interesting worldline is that which is isolated by an event. An isolated path on $T$ describes a worldline which eventually stablizes and stops to split further. This corresponds to the following notion.

\begin{defn}
Let $x$ be an event and let $T$ be a tree corresponding to the future light cone of $x$. A worldline $p$ of $x$ is called {\em eventually deterministic} if $p$ defines an isolated path on $T$.
\end{defn}

Now we will make use of the following results.

\begin{thm}\label{thm:theoremtrees}
\begin{enumerate}
    \item [(i)] Any isolated member of an effectively closed set is computable.
    \item[(ii)] An effectively closed set is called {\em special} if it does not contain a computable member. Every special effectively closed set has $2^{\aleph_0}$ many members. By contrapositive, if an effectively closed set is countable, then it contains a computable member.
    \item[(iii)] Any countable closed subset of Cantor space contains an isolated member.
\end{enumerate}
\end{thm}

\begin{prf}
\begin{enumerate}
    \item [(i)] Let $T$ be a computable tree and let $A$ be a path on $T$. Assume that $A\in[T]$ isolated. Then there is a string $\sigma\subseteq A$ such that every path above $\sigma$ is finite except $A$ (see Figure \ref{fig:isolated}). To compute the path $A$, we need to define the value of $A(n)$, that is, the $n$-th bit of the characteristic sequence of $A$. This is done by computably defining each next bit of $A$, in other words, define $A\upharpoonright n$ for each $n\in\mathbb{N}$, on $T$. We use K\"{o}nig's lemma by induction on $n$. That is, for any $n>|\sigma|$, there exists a unique $\tau\supseteq\sigma$ of length $n$ such that there is an infinite extension above $\tau$ in $T$. To compute $A\upharpoonright n$ for $n>|\sigma|$, we find $m\geq n$ such that exactly one $\tau\supseteq\sigma$ of length $n$ has an extension of length $m$ in $T$. Then, we let $A\upharpoonright n = \tau$. By the compactness of Cantor space (which is ensured by K\"{o}nig's lemma), since we compute every finite initial segment of $A$, we can compute the path $A$ on $T$.
    
    \begin{figure}[ht]
    \begin{center}
    \includegraphics{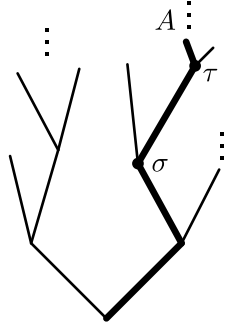}
    \caption{If a path $A$ (shown in bold lines) is isolated on a finitely branching tree $T$, then there exists some string $\sigma\subseteq A$ in $T$ such that every path above $\sigma$ is finite except $A$. To compute $A$ we inductively select a string extending $\sigma$ (such as $\tau$) which has an infinite extension in $T$.}
    \label{fig:isolated}
    \end{center}
    \end{figure}   
    
    \item[(ii)] By (i), since there is no computable member in $[T]$, every member of $T$ splits. Therefore the number of elements of $[T]$ is $2^{\aleph_0}$.
    \item[(iii)] Let $[T]$ be a closed set for a tree $T$. If $[T]$ has no isolated points, then $T$ is perfect and thus $[T$] is uncountable by Proposition \ref{prop:perfect}.\qed
\end{enumerate}
\end{prf}

One consequence of this theorem concerns finite $\Pi^0_1$ classes.

\begin{col}\label{col:corollary}
Let $\mathcal{P}$ be a finite $\Pi^0_1$ class. Then every member of $\mathcal{P}$ is computable.
\end{col}
\begin{prf}
We show that if $\mathcal{P}=[T]$ is a finite $\Pi^0_1$ class for some computable tree $T$, then every path on $T$ is isolated, and so by Theorem \ref{thm:theoremtrees}(i) every element of $[T]$ is computable. Suppose for a contradiction that there exists a non-isolated path $X$ in $[T]$. Then there must exist some $\sigma\subseteq X$ in $T$ such that $[T_\sigma]$ is infinite. Since $[T_\sigma]\subseteq [T]$, it follows that $[T]$ must be infinite. A contradiction.\qed
\end{prf}

The rest is just translating the results into the physical domain. We can argue for example that given any event $x$, if all worldlines extending $x$ are eventually deterministic, then there are countably many possible worldlines with respect to $x$. To see this, consider the future light cone of $x$, which is described by a tree $T$. If every member of $[T]$ is eventually deterministic---and so all isolated---then $[T]$ defines a discrete set. Thus $[T]$ is necessarily countable.

It also follows from Theorem \ref{thm:theoremtrees} that if there are countably many future worldlines of an event $x$, then at least one of these worldlines is decidable. 

On any tree $T\subseteq n^\mathbb{N}$, the number of finite paths can only be countable. This can be stated in terms of events and worldlines as follows.
\begin{col}\label{col:corollary2}
In the future of any event, there can only be countably many terminating worldlines.
\end{col}
\begin{prf}
    Let $x$ be an event, but suppose that the set of terminating future worldlines of $x$ is uncountable. Since no finite path can be a limit point, the set of paths that terminate can only define a discrete and countable set.\qed
\end{prf}

We also claim that given an event $x$, if every future worldline of $x$ terminates, then the collection of future worldlines of $x$ is finite. This is because of the fact that if a worldline terminates, then it cannot constitute an infinite path, hence by the contrapositive of K\"{o}nig's lemma, the set of future events of $x$ is finite. Therefore, the set of future worldlines of $x$ must be finite given the hypothesis that every future worldline of $x$ terminates. Another proof of this fact relies on a computability theoretic approach. Any terminating future worldline of an event $x$ is a finite path which just happens to extend $x$. A finite path defines a finite set. Any finite set is computable. But there can only be countably computable sets since there are only countably many computable subsets of $\mathbb{N}$. This is because there are only countably many algorithms (or Turing machine programs) that can be written in a formal language. 

One final thing before we conclude this section is to ask whether our analysis can be made on the many-worlds interpretation where the wave-function is always evolving deterministically and thus there are no probabilities. In this case, under the assumption that spacetime is discretized, the entire analysis reduces to answering the question whether there are countably or uncountably infinite orthogonal bases for this singular wave-function that encompasses all of the many-worlds. Such an investigation is out of the scope of this particular work, however, we invite interested colleagues to perform that study, which should yield qualitative differences between formulations of many-world interpretation where such a singular wave-function has countably infinite orthogonal bases in comparison to uncountably infinite ones.

\subsection{Translation between mathematical and physical contexts} \label{sec:translation}

In this part we elaborate on some of the assumptions given in Section \ref{sec:basicassumptions} and then argue how the future light cone of an event is represented by a tree structure. In this representation, worldlines of infinite length form members of $\Pi^0_1$ classes (or they form paths on the corresponding computable tree). Hence, one can treat strings in a computable tree as spacetime events in light cones, and treat worldlines as paths on the tree. Collapsing strings in a perfect tree into the elements of the set may be thought of as applying a form of {\em coarse graining} of the causal set.

We now give some elaborations of the assumptions provided in Section \ref{sec:basicassumptions} that need further explanation to complete the discussion on the representation of physical elements.

\vspace{0.5cm}

\noindent {\bf Elaboration of II(d):}

\begin{figure}[ht]
    \begin{center}
    \includegraphics{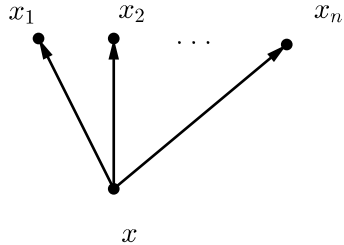}
    \caption{A split defines alternate worldlines.}
    \label{fig:split}
    \end{center}
    \end{figure}

Given an event $x$, a split of at least two immediate effects of $x$ defines incomparable spacetime events (see Figure \ref{fig:split}). In the MWI, thus, a split is usually considered as a way of defining distinct possible worlds.

\begin{figure}[ht]
    \begin{center}
    \includegraphics[scale=1.3]{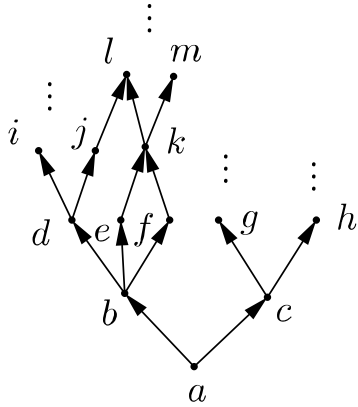}
    \caption{An example of a split-merge region between the events $b$ and $l$, and also between $b$ and $k$.}
    \label{fig:causality}
    \end{center}
    \end{figure}

Hence the assumption given in II(d) ensures the existence of alternate/possible worldlines relative to any event. One may ask what would happen if there were mergings in the light cone of an event. It may be possible that two incomparable events may later merge into a single event and continue along the same worldline (see Figure \ref{fig:causality}). Even if separate worldlines merge into a single worldline at some point in the future---in which case the structure of causal sets could be represented by a directed acyclic graph---in our case one can just use trees to represent light cones. Furthermore, splitting and merging events will not have any effect on the cardinality of worldlines. The reason is the following. The number of paths from the beginning of the split to the merging region always is finite. Let us take Figure \ref{fig:causality} as an illustration. A split occurs at event $b$ and a merging occurs at event $l$. Now there can only be finitely many events that are in the past of $l$ {\em and} the future of $b$. That is, there are finitely many $z$ such that $b\leq z\leq l$. We shall call the collection of such $z$'s a {\em split-merge region}. The number of paths that start with $b$ and end up with $l$ is clearly finite. This then has no effect on the number of worldlines in the future of $b$ if the future of $b$ contains infinitely many paths.
    
\vspace{0.5cm}

\noindent {\bf Elaboration of II(e):} We think of {\em event} as an indivisible (or atomic) entity in our discrete spacetime conception. By I(c), since the observable universe is finite and there are finitely many entities in the universe, there can only be finitely many immediate effects of an event. This is due to that there are finitely many positions a particle can be in at each next time click, where the number of positions is bounded by the size of the observable universe and the speed of light.\footnote{This discussion can possibly be extended to show that under Lorentz transformation, there exist a suitable pairing function that can translate the order of events/branches between different reference frames while maintaining the overall causal structure of forward and backward branching trees, however, we did not delve further into this line of inquiry.} This means that there are finitely many possible effects that can emerge from every event. At each next time click of spacetime evolution, the number of immediate successors of any event is bounded by a natural number. The fact that every event is immediately proceeded by a finite number of future events also guarantees that the future light cone of an event (and similarly the past light cone, if these two light cones are thought to have the symmetry of each other) can be represented by a finitely branching tree. The elements of the tree consists of spacetime events partially ordered by causation. Mathematically speaking, the elements are strings which ordered by the string extension relation $\subseteq$. So given any event $x$, the future (and the past) light cone of $x$ can be thought of as a finitely branching tree of strings. Since the universe is expanding indefinitely, the tree is potentially infinite.

\vspace{0.5cm}
\noindent{\bf Elaboration of II(f):} In causal sets, usually the anti-symmetric property is assumed automatically, and this is also what we adopt as well. However, one may ask what follows if one chooses not to add the anti-symmetric property as an axiom. The anti-symmetric property, it seems, ensures that the light cone of an event does not contain any cycles. In fact this corresponds to the concept of closed timelike curves in the causal structure. If $x\neq y$ whenever $x\leq y$ and $y\leq x$, then this means that the spacetime region that lays in-between the events $x$ and $y$ is a closed timelike curve. So the anti-symmetric property fails iff there exists a closed timelike curve. We leave it as an open question that how many possible worldlines there would be if one decided to reject the anti-symmetric property. Note that the representation of light cones changes in case one assumes the existence of closed timelike curves. 

\vspace{0.5cm}
\noindent{\bf Elaboration of II(g):} It follows from I(b), II(c), II(e) and from the previous paragraph that given any event $x$, the collection of all future events from $x$ forms a subset of $n^{<\mathbb{N}}$. II(c) ensures that if $y$ is an immediate successor of $x$, then there is no spacetime event $z$ satisfying that $x<z<y$. So given the future light cone of $x$ as a tree, all nodes in the tree are separated due to this condition. Thus, II(c) allows us to use the notion of {\em immediate effect}. And so whenever $y$ is an immediate effect of $x$, this can be simply shown by a directed branch from $x$ to $y$---like a Hasse diagram---where $x$ is put just below $y$ in the tree. Combining this with II(e) and I(d), the future light cone of an event $x$ forms a collection of events which are the immediate effects of $x$, the immediate effects of effects of $x$, and so on. This defines a subset of $n^{<\mathbb{N}}$ for some natural number $n$ since there are only finitely many immediate effects of $x$. 
\vspace{0.5cm}

\noindent{\bf Elaboration of II(h):} Since we work with cardinal arithmetic and we are interested in the number of worldlines relative to a given event $x$, we may imagine the past light cone has the symmetry of future light cone. But we need to look at two possibilities where they might differ from each other.

Possibility 1: The universe has no beginning. Assuming that the past of $x$ is eternal, future and past light cones of $x$ can be taken as two separate tree structures where one is just an inverted version of the other. So the same structure holds for both past and future light cones of an event $x$. That is, they are both infinite subsets of $n^{<\mathbb{N}}$.

Possibility 2: The universe has a beginning. In this case, the number of events that end up with $x$ is finite.

Assuming that the universe is indefinitely expanding, the number of future events of $x$ is greater than or equal to the number events in the past of $x$. So this means that, in the worst case, we can just work with the future light cone of $x$ to calculate the cardinality of the set of worldlines (past and future in complete) relative to $x$. The dominant factor without loss of generality can thus be assumed to come from the number of {\em future} worldlines of $x$.
\vspace{0.5cm}

\noindent{\bf Elaboration of III(a):}
The definition given in III(a) guarantees that a worldline of a given event $x$ is a sequence of successive events in the past and the future of $x$. A future worldline of $x$ is just a path that lies on the future light cone of $x$. Since we argued in the elaboration of II(h) that it is sufficient to work with the future light cone, it is worth noting that the future light cone of $x$ is simply a subset of $n^{<\mathbb{N}}$ as argued in the elaboration of II(g). But due to the fact that the future light cone is represented by a finitely branching tree $T$ of strings, where $x$ is the root of the tree and every successor of $x$ is an effect of $x$, the branches that lie on $T$ give us the future worldlines of $x$, that is, possible worldlines that emerge from $x$.
\vspace{0.5cm}

We have not quite argued about how to form a computable tree of strings from events. Let $x$ be a spacetime event. The future light cone of $x$ can be formed as a tree $T\subseteq n^{<\mathbb{N}}$ for some natural number $n$ inductively as follows: We let $x$ be the empty string, hence the root of the tree. Let $\sigma\in T$ represents an event $y$. Every immediate effect $z_i$ of $y$ is defined by $\sigma*i$ in $T$. 

The tree defined by the effects of an event $x$ constitutes a computable tree. This is in fact trivial since we really take the full tree $n^{<\mathbb{N}}$ to begin with, where $n$ is bounded by the spacetime region that a particle can travel from $x$ with a speed of light. So the tree $n^{<\mathbb{N}}$ consists of all immediate effects of $x$, all immediate effects of immediate effects of $x$, and so on. The future light cone of $x$ is then always computable and it basically corresponds to the full $n$-ary tree. By I(e), the immediate effects of any given event $x$ is bounded by the spacetime region through which light can travel from $x$ in least possible time tick. It is clear that the number of infinite paths on a full $n$-ary tree is uncountable since it is perfect (see Proposition \ref{prop:perfect}).\footnote{In \cite{KleitmanRothschild1975}, it was shown that as time tick $t$ increases, the number of possibilites grows exponentially. In \cite{AhmedRideout2010}, it was shown that every causal set dynamics typically yields an exponentially expanding universe. The way we define our tree and how it grows explonentially is compatible with this observation.}

Given a spacetime event $x$, it should now be clear that the future light cone of $x$ is a full $n$-ary tree. The class of all infinite paths on this tree gives us an effectively closed set. So now properties about members of effectively closed sets and the topology of Cantor space that are given here can be interpreted as properties of light cones, worldlines and events in the way we described. Each future worldline is a path on the tree of effects of $x$. 

\section{Discussion and Conclusion}

In this work we explored the relationship between the many-worlds interpretation of quantum mechanics and the cardinality of the set of possible worldlines in different circumstances under the given interpretation. We started by defining what an event is with respect to the given interpretation of spacetime, and showed how worldlines could be constructed based on this. We then associated worldlines with the paths on computable trees, and the collection of infinite paths on these trees with effectively closed subsets of Cantor space, (in fact, subsets of $n^\mathbb{N}$). Using some known topological properties of the Cantor space and computability, under the assumptions we made with respect to the nature of spacetime (i.e., that being discrete and etc.), we derived some results through finitely branching trees containing all possible worldlines as infinite paths, irrespective of whether there is an initial event (like the Big Bang) or working on an eternal universe model. Our analysis revealed the following results: In a {\em full-blown} dynamics where every event yields every possible immediate effect, in the limit we end up having uncountably many future worldlines with respect to any event. However, if all worldlines extending $x$ are `eventually deterministic', in the sense that there exists an event above which there is no more splitting, then there are exactly $\aleph_0$ many possible futures of $x$. Furthermore, if there are countably many worldlines with respect to $x$, then at least one of them must be decidable in the sense that there is an effective method, i.e. an algorithm, which determines whether or not any given event belongs to that worldline. We also pointed out that the number of terminating worldlines can only be countable, and so in a full blown dynamics the number of non-terminating worldlines is uncountable.

The summary and discussion above can also be understood in the following manner. In the MWI of quantum mechanics, there exist all the possible worlds with all possible configurations. Universes with no beginning, universes with singular beginnings, universes with no ending, or multiple endings, eternal universes with no ending, universes with no beginning but that will have an ending, and so on. In our analysis, we identified a relationship which states that, under certain discrete spacetime assumptions, given an event $x$, if all worldlines extending that event lead to eventually deterministic futures, then at least one of them must be decidable. The converse of this conditional statement may not always hold. Hence, our work actually identifies that for certain types of dynamics---where events in a future light cone lead to eventually deterministic futures---there is a decidable worldline and that there can only countably many eventually deterministic worldlines. We believe that, the property of decidability is an important distinction for an interpretation of quantum mechanics, for that models which are unable to produce decidable worldlines should not be considered as physically feasible. Moreover, decidability features of causal sets is not something new. A related study regarding this can also be found in \cite{Bolognesi2011}, where the author introduces how to generate algorithmic causal sets using models of computations and to perform computer simulations on generated sets. This is possible only if the set of future worldlines of an event contains an eventually determnistic worldline, or that the set is countable.

We consider this work as a part of the bigger goal of creating and using rigorous mathematical tools and deductive methods to apply on the problem of distinguishing between different interpretations of quantum mechanics. We find it particularly interesting the apparent interplay between causal sets, which are described by acyclic directed graphs, and computable trees or effectively closed subsets of Cantor space described in a similar manner. Some mathematical theorems about effectively closed sets or the topology of Cantor space in general, thus, could be applied on causal set theories so as to gain insights for solving various cosmological problems of quantum theory, whose theory of gravity may be modelled by causal sets. As the field of quantum technologies progress rapidly, the need to address the still unresolved questions from the early days of quantum mechanics should be identified as a worthy endeavour, and we want to strongly encourage the readers of this work to contemplate on these issues. Formulation of clear experiments to rule out certain interpretations can only come from delving deep into the assumptions and their implications, and it is a long and arduous journey that awaits the community of researchers both in foundations of quantum mechanics and philosophy of physics.

\section*{Acknowledgement}
We would like to thank the anonymous referee for many useful suggestions which significantly improved the quality of this work. We would also like to thank David D. Reid, and \"{O}zlem Salehi for their valuable feedback and comments.


\end{document}